\documentclass[longauth,twocolumn,traditabstract]{aa}
\usepackage[table,usenames,dvipsnames]{xcolor}
\usepackage{enumerate}
\usepackage{ifthen}
\usepackage{multirow}
\usepackage{upgreek}
\usepackage{overpic}
\usepackage{graphicx,amsmath}
\usepackage{txfonts}
\usepackage[breaklinks, colorlinks, citecolor=blue, colorlinks=true, linkcolor=blue]{hyperref}
\usepackage{natbib}
\bibpunct{(}{)}{;}{a}{}{,} 
\usepackage{pdfcolmk} 
\usepackage[table]{xcolor}
\usepackage{colortab}
\usepackage{soul}
\usepackage[normalem]{ulem}
\usepackage{amssymb}

\pdfmapfile{+txfonts.map}

\def\setsymbol#1#2{\expandafter\def\csname #1\endcsname{#2}}
\def\getsymbol#1{\csname #1\endcsname}

\def\Planck{\textit{Planck}}

\def\all2013resultspapers{\nocite{planck2013-p01, planck2013-p02, planck2013-p02a, planck2013-p02d, planck2013-p02b, planck2013-p03, planck2013-p03c, planck2013-p03f, planck2013-p03d, planck2013-p03e, planck2013-p01a, planck2013-p06, planck2013-p03a, planck2013-pip88, planck2013-p08, planck2013-p11, planck2013-p12, planck2013-p13, planck2013-p14, planck2013-p15, planck2013-p05b, planck2013-p17, planck2013-p09, planck2013-p09a, planck2013-p20, planck2013-p19, planck2013-pipaberration, planck2013-p05, planck2013-p05a, planck2013-pip56, planck2013-p06b, planck2013-p01a}}

\newbox\tablebox    \newdimen\tablewidth
\def\leaderfil{\leaders\hbox to 5pt{\hss.\hss}\hfil}
\def\endPlancktable{\tablewidth=\columnwidth 
    $$\hss\copy\tablebox\hss$$
    \vskip-\lastskip\vskip -2pt}

\def\tablenote#1 #2\par{\begingroup \parindent=0.8em
    \abovedisplayshortskip=0pt\belowdisplayshortskip=0pt
    \noindent
    $$\hss\vbox{\hsize\tablewidth \hangindent=\parindent \hangafter=1 \noindent
    \hbox to \parindent{$^#1$\hss}\strut#2\strut\par}\hss$$
    \endgroup}
\def\doubleline{\vskip 3pt\hrule \vskip 1.5pt \hrule \vskip 5pt}

\def\L2{\ifmmode L_2\else $L_2$\fi}

\def\DeltaT{\ifmmode \Delta T\else $\Delta T$\fi}
\def\deltat{\ifmmode \Delta t\else $\Delta t$\fi}
\def\fknee{\ifmmode f_{\rm knee}\else $f_{\rm knee}$\fi}
\def\Fmax{\ifmmode F_{\rm max}\else $F_{\rm max}$\fi}
\def\solar{\ifmmode{\rm M}_{\mathord\odot}\else${\rm M}_{\mathord\odot}$\fi}
\def\Msolar{\ifmmode{\rm M}_{\mathord\odot}\else${\rm M}_{\mathord\odot}$\fi}
\def\Lsolar{\ifmmode{\rm L}_{\mathord\odot}\else${\rm L}_{\mathord\odot}$\fi}
\def\inv{\ifmmode^{-1}\else$^{-1}$\fi}
\def\mo{\ifmmode^{-1}\else$^{-1}$\fi}
\def\sup#1{\ifmmode ^{\rm #1}\else $^{\rm #1}$\fi}
\def\expo#1{\ifmmode \times 10^{#1}\else $\times 10^{#1}$\fi}
\def\,{\thinspace}
\def\lsim{\mathrel{\raise .4ex\hbox{\rlap{$<$}\lower 1.2ex\hbox{$\sim$}}}}
\def\gsim{\mathrel{\raise .4ex\hbox{\rlap{$>$}\lower 1.2ex\hbox{$\sim$}}}}

\def\simprop{\mathrel{\raise .4ex\hbox{\rlap{$\propto$}\lower 1.2ex\hbox{$\sim$}}}}
\def\deg{\ifmmode^\circ\else$^\circ$\fi}
\def\pdeg{\ifmmode $\setbox0=\hbox{$^{\circ}$}\rlap{\hskip.11\wd0 .}$^{\circ}
          \else \setbox0=\hbox{$^{\circ}$}\rlap{\hskip.11\wd0 .}$^{\circ}$\fi}
\def\arcs{\ifmmode {^{\scriptstyle\prime\prime}}
          \else $^{\scriptstyle\prime\prime}$\fi}
\def\arcm{\ifmmode {^{\scriptstyle\prime}}
          \else $^{\scriptstyle\prime}$\fi}
\newdimen\sa  \newdimen\sb
\def\parcs{\sa=.07em \sb=.03em
     \ifmmode \hbox{\rlap{.}}^{\scriptstyle\prime\kern -\sb\prime}\hbox{\kern -\sa}
     \else \rlap{.}$^{\scriptstyle\prime\kern -\sb\prime}$\kern -\sa\fi}
\def\parcm{\sa=.08em \sb=.03em
     \ifmmode \hbox{\rlap{.}\kern\sa}^{\scriptstyle\prime}\hbox{\kern-\sb}
     \else \rlap{.}\kern\sa$^{\scriptstyle\prime}$\kern-\sb\fi}
\def\ra[#1 #2 #3.#4]{#1\sup{h}#2\sup{m}#3\sup{s}\llap.#4}
\def\dec[#1 #2 #3.#4]{#1\deg#2\arcm#3\arcs\llap.#4}
\def\deco[#1 #2 #3]{#1\deg#2\arcm#3\arcs}
\def\rra[#1 #2]{#1\sup{h}#2\sup{m}}

\def\dots{\relax\ifmmode \ldots\else $\ldots$\fi}
\def\WHzsr{\ifmmode $W\,Hz\mo\,sr\mo$\else W\,Hz\mo\,sr\mo\fi}
\def\mHz{\ifmmode $\,mHz$\else \,mHz\fi}
\def\GHz{\ifmmode $\,GHz$\else \,GHz\fi}
\def\mKs{\ifmmode $\,mK\,s$^{1/2}\else \,mK\,s$^{1/2}$\fi}
\def\muKs{\ifmmode \,\mu$K\,s$^{1/2}\else \,$\mu$K\,s$^{1/2}$\fi}
\def\muKRJs{\ifmmode \,\mu$K$_{\rm RJ}$\,s$^{1/2}\else \,$\mu$K$_{\rm RJ}$\,s$^{1/2}$\fi}
\def\muKHz{\ifmmode \,\mu$K\,Hz$^{-1/2}\else \,$\mu$K\,Hz$^{-1/2}$\fi}
\def\MJysr{\ifmmode \,$MJy\,sr\mo$\else \,MJy\,sr\mo\fi}
\def\MJysrmK{\ifmmode \,$MJy\,sr\mo$\,mK$_{\rm CMB}\mo\else \,MJy\,sr\mo\,mK$_{\rm CMB}\mo$\fi}
\def\microns{\ifmmode \,\mu$m$\else \,$\mu$m\fi}

\def\muK{\ifmmode \,\mu$K$\else \,$\mu$\hbox{K}\fi}
\def\microK{\ifmmode \,\mu$K$\else \,$\mu$\hbox{K}\fi}
\def\muW{\ifmmode \,\mu$W$\else \,$\mu$\hbox{W}\fi}
\def\kms{\ifmmode $\,km\,s$^{-1}\else \,km\,s$^{-1}$\fi}
\def\kmsMpc{\ifmmode $\,\kms\,Mpc\mo$\else \,\kms\,Mpc\mo\fi}
\providecommand{\sorthelp}[1]{}

\newcommand{\be}{\begin{equation}}
\newcommand{\ee}{\end{equation}}

\newcommand{\threej}[6]{\left(
    \begin{array}{ccc}
        \! #1\! & #2\!  & #3\!  \\
        \! #4\! & #5\!  & #6\!
      \end{array}
    \right)}
    
\newcommand{\hatn}{\vec{\hat{n\,}}}

\newcommand{\aberration}{aberration}
\newcommand{\modulation}{modulation}

\newcommand{\perpa}{\perp}
\newcommand{\perpb}{\times}

\newcommand{\fnu}{b_{\nu}}

\newcommand{\betanu}{\beta_{\nu}}

\begin{document}

\title{\Planck\ 2013 results. XXVII. Doppler boosting of the CMB: \\
Eppur si muove\thanks{``And yet it
moves,'' the phrase popularly attributed to Galileo Galilei after being
forced to recant his view that the Earth goes around the Sun.}} 
\author{\small
Planck Collaboration:
N.~Aghanim\inst{56}
\and
C.~Armitage-Caplan\inst{85}
\and
M.~Arnaud\inst{69}
\and
M.~Ashdown\inst{66, 6}
\and
F.~Atrio-Barandela\inst{16}
\and
J.~Aumont\inst{56}
\and
C.~Baccigalupi\inst{79}
\and
A.~J.~Banday\inst{87, 9}
\and
R.~B.~Barreiro\inst{63}
\and
J.~G.~Bartlett\inst{1, 64}
% ...
}
\institute{\small
APC, AstroParticule et Cosmologie, Universit\'{e} Paris Diderot, CNRS/IN2P3, CEA/lrfu, Observatoire de Paris, Sorbonne Paris Cit\'{e}, 10, rue Alice Domon et L\'{e}onie Duquet, 75205 Paris Cedex 13, France\\
\and
Aalto University Mets\"{a}hovi Radio Observatory and Dept of Radio Science and Engineering, P.O. Box 13000, FI-00076 AALTO, Finland\\
\and
African Institute for Mathematical Sciences, 6-8 Melrose Road, Muizenberg, Cape Town, South Africa\\
\and
Agenzia Spaziale Italiana Science Data Center, Via del Politecnico snc, 00133, Roma, Italy\\
\and
Agenzia Spaziale Italiana, Viale Liegi 26, Roma, Italy\\
\and
Astrophysics Group, Cavendish Laboratory, University of Cambridge, J J Thomson Avenue, Cambridge CB3 0HE, U.K.\\
\and
Astrophysics \& Cosmology Research Unit, School of Mathematics, Statistics \& Computer Science, University of KwaZulu-Natal, Westville Campus, Private Bag X54001, Durban 4000, South Africa\\
\and
CITA, University of Toronto, 60 St. George St., Toronto, ON M5S 3H8, Canada\\
\and
CNRS, IRAP, 9 Av. colonel Roche, BP 44346, F-31028 Toulouse cedex 4, France\\
\and
California Institute of Technology, Pasadena, California, U.S.A.\\
\and
Centre for Theoretical Cosmology, DAMTP, University of Cambridge, Wilberforce Road, Cambridge CB3 0WA, U.K.\\
\and
Computational Cosmology Center, Lawrence Berkeley National Laboratory, Berkeley, California, U.S.A.\\
\and
DSM/Irfu/SPP, CEA-Saclay, F-91191 Gif-sur-Yvette Cedex, France\\
\and
DTU Space, National Space Institute, Technical University of Denmark, Elektrovej 327, DK-2800 Kgs. Lyngby, Denmark\\
\and
D\'{e}partement de Physique Th\'{e}orique, Universit\'{e} de Gen\`{e}ve, 24, Quai E. Ansermet,1211 Gen\`{e}ve 4, Switzerland\\
\and
Departamento de F\'{\i}sica Fundamental, Facultad de Ciencias, Universidad de Salamanca, 37008 Salamanca, Spain\\
\and
Departamento de F\'{\i}sica, Universidad de Oviedo, Avda. Calvo Sotelo s/n, Oviedo, Spain\\
\and
Department of Astrophysics/IMAPP, Radboud University Nijmegen, P.O. Box 9010, 6500 GL Nijmegen, The Netherlands\\
\and
Department of Electrical Engineering and Computer Sciences, University of California, Berkeley, California, U.S.A.\\
\and
Department of Physics \& Astronomy, University of British Columbia, 6224 Agricultural Road, Vancouver, British Columbia, Canada\\
\and
Department of Physics and Astronomy, Dana and David Dornsife College of Letter, Arts and Sciences, University of Southern California, Los Angeles, CA 90089, U.S.A.\\
\and
Department of Physics and Astronomy, University College London, London WC1E 6BT, U.K.\\
\and
Department of Physics and Astronomy, University of Sussex, Brighton BN1 9QH, U.K.\\
\and
Department of Physics, Florida State University, Keen Physics Building, 77 Chieftan Way, Tallahassee, Florida, U.S.A.\\
\and
Department of Physics, Gustaf H\"{a}llstr\"{o}min katu 2a, University of Helsinki, Helsinki, Finland\\
\and
Department of Physics, Princeton University, Princeton, New Jersey, U.S.A.\\
\and
Department of Physics, University of California, Berkeley, California, U.S.A.\\
\and
Department of Physics, University of California, One Shields Avenue, Davis, California, U.S.A.\\
\and
Department of Physics, University of California, Santa Barbara, California, U.S.A.\\
\and
Department of Physics, University of Illinois at Urbana-Champaign, 1110 West Green Street, Urbana, Illinois, U.S.A.\\
\and
Dipartimento di Fisica e Astronomia G. Galilei, Universit\`{a} degli Studi di Padova, via Marzolo 8, 35131 Padova, Italy\\
\and
Dipartimento di Fisica e Scienze della Terra, Universit\`{a} di Ferrara, Via Saragat 1, 44122 Ferrara, Italy\\
\and
Dipartimento di Fisica, Universit\`{a} La Sapienza, P. le A. Moro 2, Roma, Italy\\
\and
Dipartimento di Fisica, Universit\`{a} degli Studi di Milano, Via Celoria, 16, Milano, Italy\\
\and
Dipartimento di Fisica, Universit\`{a} degli Studi di Trieste, via A. Valerio 2, Trieste, Italy\\
\and
Dipartimento di Fisica, Universit\`{a} di Roma Tor Vergata, Via della Ricerca Scientifica, 1, Roma, Italy\\
\and
Discovery Center, Niels Bohr Institute, Blegdamsvej 17, Copenhagen, Denmark\\
\and
Dpto. Astrof\'{i}sica, Universidad de La Laguna (ULL), E-38206 La Laguna, Tenerife, Spain\\
\and
European Space Agency, ESAC, Planck Science Office, Camino bajo del Castillo, s/n, Urbanizaci\'{o}n Villafranca del Castillo, Villanueva de la Ca\~{n}ada, Madrid, Spain\\
\and
European Space Agency, ESTEC, Keplerlaan 1, 2201 AZ Noordwijk, The Netherlands\\
\and
Helsinki Institute of Physics, Gustaf H\"{a}llstr\"{o}min katu 2, University of Helsinki, Helsinki, Finland\\
\and
INAF - Osservatorio Astronomico di Padova, Vicolo dell'Osservatorio 5, Padova, Italy\\
\and
INAF - Osservatorio Astronomico di Roma, via di Frascati 33, Monte Porzio Catone, Italy\\
\and
INAF - Osservatorio Astronomico di Trieste, Via G.B. Tiepolo 11, Trieste, Italy\\
\and
INAF Istituto di Radioastronomia, Via P. Gobetti 101, 40129 Bologna, Italy\\
\and
INAF/IASF Bologna, Via Gobetti 101, Bologna, Italy\\
\and
INAF/IASF Milano, Via E. Bassini 15, Milano, Italy\\
\and
INFN, Sezione di Bologna, Via Irnerio 46, I-40126, Bologna, Italy\\
\and
INFN, Sezione di Roma 1, Universit\`{a} di Roma Sapienza, Piazzale Aldo Moro 2, 00185, Roma, Italy\\
\and
INFN/National Institute for Nuclear Physics, Via Valerio 2, I-34127 Trieste, Italy\\
\and
IPAG: Institut de Plan\'{e}tologie et d'Astrophysique de Grenoble, Universit\'{e} Joseph Fourier, Grenoble 1 / CNRS-INSU, UMR 5274, Grenoble, F-38041, France\\
\and
ISDC Data Centre for Astrophysics, University of Geneva, ch. d'Ecogia 16, Versoix, Switzerland\\
\and
IUCAA, Post Bag 4, Ganeshkhind, Pune University Campus, Pune 411 007, India\\
\and
Imperial College London, Astrophysics group, Blackett Laboratory, Prince Consort Road, London, SW7 2AZ, U.K.\\
\and
Infrared Processing and Analysis Center, California Institute of Technology, Pasadena, CA 91125, U.S.A.\\
\and
Institut d'Astrophysique Spatiale, CNRS (UMR8617) Universit\'{e} Paris-Sud 11, B\^{a}timent 121, Orsay, France\\
\and
Institut d'Astrophysique de Paris, CNRS (UMR7095), 98 bis Boulevard Arago, F-75014, Paris, France\\
\and
Institute for Space Sciences, Bucharest-Magurale, Romania\\
\and
Institute of Astronomy and Astrophysics, Academia Sinica, Taipei, Taiwan\\
\and
Institute of Astronomy, University of Cambridge, Madingley Road, Cambridge CB3 0HA, U.K.\\
\and
Institute of Theoretical Astrophysics, University of Oslo, Blindern, Oslo, Norway\\
\and
Instituto de Astrof\'{\i}sica de Canarias, C/V\'{\i}a L\'{a}ctea s/n, La Laguna, Tenerife, Spain\\
\and
Instituto de F\'{\i}sica de Cantabria (CSIC-Universidad de Cantabria), Avda. de los Castros s/n, Santander, Spain\\
\and
Jet Propulsion Laboratory, California Institute of Technology, 4800 Oak Grove Drive, Pasadena, California, U.S.A.\\
\and
Jodrell Bank Centre for Astrophysics, Alan Turing Building, School of Physics and Astronomy, The University of Manchester, Oxford Road, Manchester, M13 9PL, U.K.\\
\and
Kavli Institute for Cosmology Cambridge, Madingley Road, Cambridge, CB3 0HA, U.K.\\
\and
LAL, Universit\'{e} Paris-Sud, CNRS/IN2P3, Orsay, France\\
\and
LERMA, CNRS, Observatoire de Paris, 61 Avenue de l'Observatoire, Paris, France\\
\and
Laboratoire AIM, IRFU/Service d'Astrophysique - CEA/DSM - CNRS - Universit\'{e} Paris Diderot, B\^{a}t. 709, CEA-Saclay, F-91191 Gif-sur-Yvette Cedex, France\\
\and
Laboratoire Traitement et Communication de l'Information, CNRS (UMR 5141) and T\'{e}l\'{e}com ParisTech, 46 rue Barrault F-75634 Paris Cedex 13, France\\
\and
Laboratoire de Physique Subatomique et de Cosmologie, Universit\'{e} Joseph Fourier Grenoble I, CNRS/IN2P3, Institut National Polytechnique de Grenoble, 53 rue des Martyrs, 38026 Grenoble cedex, France\\
\and
Laboratoire de Physique Th\'{e}orique, Universit\'{e} Paris-Sud 11 \& CNRS, B\^{a}timent 210, 91405 Orsay, France\\
\and
Lawrence Berkeley National Laboratory, Berkeley, California, U.S.A.\\
\and
Max-Planck-Institut f\"{u}r Astrophysik, Karl-Schwarzschild-Str. 1, 85741 Garching, Germany\\
\and
McGill Physics, Ernest Rutherford Physics Building, McGill University, 3600 rue University, Montr\'{e}al, QC, H3A 2T8, Canada\\
\and
Niels Bohr Institute, Blegdamsvej 17, Copenhagen, Denmark\\
\and
Observational Cosmology, Mail Stop 367-17, California Institute of Technology, Pasadena, CA, 91125, U.S.A.\\
\and
Optical Science Laboratory, University College London, Gower Street, London, U.K.\\
\and
SISSA, Astrophysics Sector, via Bonomea 265, 34136, Trieste, Italy\\
\and
School of Physics and Astronomy, Cardiff University, Queens Buildings, The Parade, Cardiff, CF24 3AA, U.K.\\
\and
School of Physics and Astronomy, University of Nottingham, Nottingham NG7 2RD, U.K.\\
\and
Space Research Institute (IKI), Russian Academy of Sciences, Profsoyuznaya Str, 84/32, Moscow, 117997, Russia\\
\and
Space Sciences Laboratory, University of California, Berkeley, California, U.S.A.\\
\and
Stanford University, Dept of Physics, Varian Physics Bldg, 382 Via Pueblo Mall, Stanford, California, U.S.A.\\
\and
Sub-Department of Astrophysics, University of Oxford, Keble Road, Oxford OX1 3RH, U.K.\\
\and
UPMC Univ Paris 06, UMR7095, 98 bis Boulevard Arago, F-75014, Paris, France\\
\and
Universit\'{e} de Toulouse, UPS-OMP, IRAP, F-31028 Toulouse cedex 4, France\\
% ...
}

   \abstract{Our velocity relative to the rest frame of the cosmic
   microwave background (CMB)
   generates a dipole temperature anisotropy on the sky
   which has been well measured for more than 30 years, and has
   an accepted amplitude of $v/c = 1.23 \times 10^{-3}$, or $v = 369\kms$.
   In addition to this signal generated by Doppler boosting of the CMB monopole,
   our motion also modulates and aberrates the CMB temperature fluctuations
   (as well as every other source of radiation at cosmological distances).
   This is an order $10^{-3}$ effect applied to fluctuations which are
   already one part in roughly $10^{5}$, so it is quite small.
   Nevertheless, it becomes detectable with the all-sky coverage,
   high angular resolution, and low noise levels of the \Planck\ satellite.
   Here we report a first measurement of this velocity signature using
   the aberration and modulation effects on the CMB temperature anisotropies,
   finding a component in the known dipole direction,
   $(l,b)=(264^\circ, 48^\circ)$, 
   of $384\kms \pm 78\kms\ ({\rm stat.}) \pm 115\kms \ ({\rm syst.})$.  
   This is a significant confirmation of the expected velocity.
   }
   \keywords{Cosmology: observations -- cosmic background radiation
   -- Reference systems -- Relativistic processes}
   
\date{A\&A: submitted, 23 March 2013, accepted 3 January 2014.}
   
\titlerunning{Doppler boosting of the CMB: Eppur si muove}
\authorrunning{Planck Collaboration}
\maketitle

\addtocounter{footnote}{3}

\section{Introduction}

This paper, one of a set associated with the 2013 release of data from the
\Planck\footnote{\Planck\ (\url{http://www.esa.int/Planck}) is a project of
the European Space Agency (ESA) with instruments provided by two scientific
consortia funded by ESA member states (in particular the lead countries France
and Italy), with contributions from NASA (USA) and telescope reflectors
provided by a collaboration between ESA and a scientific consortium led and
funded by Denmark.}
mission \citep{planck2013-p01}, presents a study of Doppler boosting effects
using the small-scale temperature fluctuations of the \Planck\ cosmic
microwave background (CMB) maps.

Observations of the relatively large amplitude CMB
temperature dipole are usually taken to indicate that our
Solar System barycentre is in motion with respect to the CMB frame 
(defined precisely below).  
Assuming that the observed temperature dipole is due entirely
to Doppler boosting of the CMB monopole, one infers a velocity
$v = (369 \pm 0.9)\,{\rm km}\,{\rm s}^{-1}$ in the direction
$(l,b)=(263\pdeg99 \pm 0\pdeg14, 48\pdeg26 \pm 0\pdeg03)$, on the
boundary of the constellations of Crater and Leo
\citep{Kogut:1993ag,Fixsen:1996nj,Hinshaw:2008kr}.

\begin{figure}[!t]
\vspace{0.15in}
\begin{center}
\begin{overpic}[width=0.9\columnwidth]{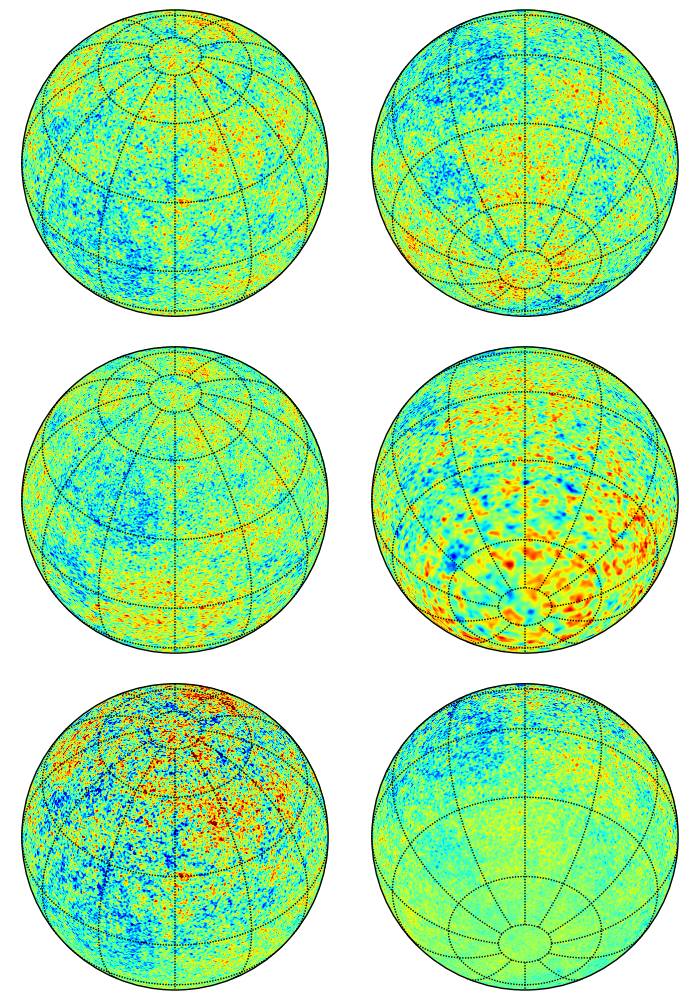}
 \put(2, 99){$(a)\ T^{\mathsc{primordial}}$}
 \put(2, 65){$(b)\ T^{\mathsc{\aberration}}$}
 \put(2, 31.5){$(c)\ T^{\mathsc{\modulation}}$} 
\end{overpic}
\caption{
Exaggerated illustration of the \aberration\ and Doppler \modulation\ effects,
in orthographic
projection, for a velocity $v = 260\,000\,{\rm km}\,{\rm s}^{-1}
 = 0.85 c$ (approximately $700$ times larger than the expected magnitude)
toward the northern pole (indicated by meridians in the upper half of each
image on the left).
The \aberration\ component of the effect shifts the apparent position of
fluctuations toward the velocity direction, while
the \modulation\  component enhances the fluctuations in the velocity direction and suppresses them in the anti-velocity direction.
\label{fig:pecvec_effects}
}
\end{center}
\end{figure}

In addition to Doppler boosting of the CMB monopole, velocity effects
also boost the order $10^{-5}$ primordial temperature fluctuations.  
There are two observable effects here, both at a
level of $\beta \equiv v/c = 1.23 \times 10^{-3}$.
First, there is a Doppler ``\modulation'' effect, which amplifies the apparent 
temperature fluctuations in the velocity direction, and reduces them in the
opposite direction.  
This is the same effect which converts a portion of the CMB monopole
into the observed dipole.  The effect on the CMB fluctuations is to
increase the amplitude of the power spectrum by approximately $0.25\%$
in the velocity direction, and decrease it correspondingly
in the anti-direction.
Second, there is an ``\aberration'' effect, in which the apparent arrival direction of
CMB photons is pushed toward the velocity direction.
This effect is small, but non-negligible.
The expected velocity induces a peak deflection of $\beta = 4\parcm2$
and a root-mean-squared (rms) deflection over the sky of $3^\prime$,
comparable to the effects of gravitational lensing by large-scale structure,
which are discussed in \cite{planck2013-p12}.  
The \aberration\ effect squashes the anisotropy pattern on one side of 
the sky and stretches it on the other, effectively changing the angular scale.
Close to the velocity direction we expect that the power spectrum of the 
temperature anisotropies, $C_\ell$, will
be shifted so that, e.g., $\ell\,{=}\,1000\to\ell\,{=}\,1001$, while
$\ell\,{=}\,1000\to\ell\,{=}\,999$ in the anti-direction.
In Fig.~\ref{fig:pecvec_effects} we plot an exaggerated illustration of the
\aberration\ and \modulation\ effects.  For completeness we should point out
that there is a third effect, a quadrupole of amplitude $\beta^2$ induced by
the dipole \citep[see][]{Kamionkowski:2002nd}.  However, extracting this
signal would require extreme levels of precision for foreground modelling at
the quadrupole scale, and we do not discuss it further.

In this paper, we will present a measurement of $\beta$, 
exploiting the distinctive statistical signatures of the \aberration\ and
\modulation\ effects on the high-$\ell$ CMB temperature anisotropies.  
In addition to our interest in making an independent measurement of the
velocity signature,
the effects which velocity generates on the CMB fluctuations provide
a source of potential bias or confusion for several aspects of the
\Planck\ data.  
In particular, velocity effects couple to measurements of:
primordial ``$\tau_{\rm NL}$''-type non-Gaussianity, 
as discussed in \cite{planck2013-p09a};
statistical anisotropy of the primordial CMB fluctuations, as discussed in
\cite{planck2013-p09}; and gravitational lensing, as discussed in
\cite{planck2013-p12}.
There are also aspects of the \Planck\ analysis for which velocity
effects are {\it believed\/} to be negligible, but only if they are
present at the expected level.
One example is measurement of $f_{\rm NL}$-type non-Gaussianity,
as discussed in \cite{Catena:2013qd}.
Another example is power spectrum estimation ---
as discussed above, velocity effects change the angular scale 
of the acoustic peaks in the CMB power spectrum.
Averaged over the full sky this effect is strongly suppressed, as the
expansion and contraction of scales on opposing hemispheres cancel
out.  However the application of a sky mask breaks this cancellation
to some extent, and can potentially be important for parameter estimation
\citep{Pereira:2010dn,Catena:2012hq}.
For the 143 and 217\,GHz analysis mask used in the fiducial \Planck\ 
CMB likelihood \citep{planck2013-p08}, the average lensing convergence
field associated with the aberration effect
(on the portion of the sky which is unmasked) has a value which is $13\%$
of its peak value, corresponding to an expected average lensing convergence 
of $\beta \times 0.13 = 1.5 \times 10^{-4}$.
This will shift the angular scale of the acoustic peaks by the same fraction,
which is degenerate with a change in the angular size of the sound horizon
at last scattering, $\theta_*$ \citep{Burles:2006xf}.
A $1.5 \times 10^{-4}$ shift in $\theta_*$ is just under $25\%$ of the 
\Planck\ uncertainty on this parameter, as reported 
in \cite{planck2013-p11} --- small enough to be neglected, though not
dramatically so.
This therefore motivates us to test that the observed aberration signal
is not significantly larger than expected.
With such a confirmation in hand, a logical next step is to correct for
these effects by a process of de-boosting the observed temperature
\citep{Notari:2011sb,Yoho:2012am}.
Indeed, an analysis of maps corrected for the modulation effect described
here is performed in 
\cite{planck2013-p09}.

Before proceeding to discuss the \aberration\ and \modulation\ effects
in more detail, we note that in addition to the overall peculiar
velocity of our Solar System with respect
to the CMB, there is an additional time-dependent velocity effect
from the orbit of \Planck\ (at L2, along with the Earth) about the Sun.
This velocity has an average amplitude of approximately
$30\,{\rm km}\,{\rm s}^{-1}$, less than one-tenth the size of the primary
velocity effect. 
The \aberration\ component of the orbital velocity (more commonly referred
to in astronomy as ``stellar aberration'') has a maximum amplitude of
$20\parcs5$ and is corrected for in the satellite pointing.
The \modulation\  effect for the orbital velocity switches signs between
each 6-month survey, and so is suppressed when using multiple surveys to make
maps (as we do here, with the nominal \Planck\ maps, based on a little more
than two surveys), and so we will not consider it further.\footnote{Note that 
in both stellar and cosmological cases, the aberration is the result of 
{\em local} velocity differences \citep{eisner1967,phipps1989}: in the former 
case, between Earth's velocity at different times of the year, and in the 
latter between the actual and CMB frames.}

\section{Aberration and modulation}

Here we will present a more quantitative description of the \aberration\
and \modulation\ effects described above.  To begin, note that, by 
construction, the {\em peculiar velocity}, $\vec\beta$, measures the 
velocity of our Solar System barycentre relative to a frame, called the 
{\em CMB frame}, in which the temperature dipole, $a_{1m}$, vanishes.  
However, in completely subtracting the dipole, this frame would not 
coincide with a suitably-defined {\em average} CMB frame, in which an 
observer would expect to see a dipole $C_1 \sim 10^{-10}$, 
given by the Sachs-Wolfe and integrated Sachs-Wolfe effects 
(see \citet{Zibin:2008} for discussion of cosmic variance in the CMB monopole 
and dipole).  The 
velocity difference between these two frames is, however, small, at the 
level of 1\% of our observed $v$.

If $T^\prime$ and $\hatn^\prime$ are the CMB temperature and direction as 
viewed in the CMB frame, then the temperature in the observed frame is 
given by the Lorentz transformation 
\citep[see, e.g.,][]{Challinor:2002zh,Sollom:2010},
\be
T(\hatn) = \frac{T^\prime(\hatn^\prime)}{\gamma(1 - \hatn\cdot\vec{\beta})},
\ee
where the observed direction $\hatn$ is given by
\be
\hatn = \frac{\hatn^\prime + [(\gamma - 1)\hatn^\prime\cdot\hat{\vec{v}}
      + \gamma\beta]\hat{\vec{v}}}{\gamma(1 + \hatn^\prime\cdot\vec{\beta})},
\ee
and $\gamma \equiv (1 - \beta^2)^{-1/2}$.  Expanding to linear order in 
$\beta$ gives
\be
T^\prime(\hatn^\prime) = T^\prime(\hatn - \nabla(\hatn\cdot\vec{\beta}))
   \equiv T_0 + \delta T^\prime(\hatn - \nabla(\hatn\cdot\vec{\beta})),
\ee
so that we can write the observed temperature fluctuations as
\be
\delta T(\hatn) = T_0\hatn\cdot\vec{\beta} + 
   \delta T^\prime(\hatn - \nabla(\hatn\cdot\vec{\beta}))
   (1 + \hatn\cdot\vec{\beta}).
\label{delTdecomp}
\ee
Here $T_0 = (2.7255 \pm 0.0006)$\,K is the CMB mean temperature 
\citep{fixsen2009}.  The first term on the right-hand side of 
Eq.~\eqref{delTdecomp} is the temperature dipole.  The remaining term 
represents the fluctuations, aberrated by deflection 
$\nabla(\hatn\cdot\vec{\beta})$ and modulated by the factor 
$(1 + \hatn\cdot\vec{\beta})$.

The \Planck\ detectors can be modelled as measuring differential 
changes in the CMB intensity at frequency $\nu$ given by
\be
I_\nu(\nu, \hatn) = \frac{2 h \nu^3}{c^2}
 \frac{1}{ \exp \left[ h \nu / k_{\rm B} T(\hatn) \right] - 1 }.
\ee
We can expand the measured intensity difference according to
\be
\delta I_\nu(\nu,\hatn) = \left.\frac{dI_\nu}{dT}\right|_{T_0} \delta T(\hatn)
   + \frac{1}{2}\left.\frac{d^2I_\nu}{dT^2}\right|_{T_0} \delta T^2(\hatn)
   + \dots\ .
\label{Inuexpn}
\ee
Substituting Eq.~\eqref{delTdecomp} and dropping terms of order $\beta^2$ and 
$(\delta T^\prime)^2$, we find
\be
\delta I_\nu(\nu,\hatn) = \left.\frac{dI_\nu}{dT}\right|_{T_0}\!
   \left[T_0\hatn\cdot\vec{\beta}
 + \delta T^\prime(\hatn^\prime)(1 + \fnu\hatn\cdot\vec{\beta})\right],
\ee
where the frequency dependent boost factor $\fnu$ is given by
\be
\fnu = \frac{\nu}{ \nu_0 } \coth\left( \frac{\nu}{2 \nu_0} \right) - 1,
\ee
with $\nu_0 \equiv k_{\rm B} T_0 / h \simeq 57\,{\rm GHz}$.  
Integrated over the \Planck\ bandpasses for
the 143 and 217\,GHz channels,  
these effective boost factors are given by
$b_{143} = 1.961 \pm 0.015$, and
$b_{217} = 3.071 \pm 0.018$,
where the uncertainties represent the scatter between the individual detector
bandpasses at each frequency.
We will approximate these boost factors simply as 
$b_{143} = 2$ and $b_{217} = 3$,
which is sufficiently accurate for the precision of our measurement.

In the mapmaking process, the fluctuations are assumed to satisfy only the 
linear term in Eq.~\eqref{Inuexpn}.  Therefore, the inferred temperature 
fluctuations will be
\be
\frac{\delta I_\nu(\nu,\hatn)}{dI_\nu/dT|_{T_0}} = T_0\hatn\cdot\vec{\beta}
 + \delta T^\prime(\hatn - \nabla(\hatn\cdot\vec{\beta}))
 (1 + \fnu\hatn\cdot\vec{\beta}).
\label{delTinfer}
\ee
Notice that, compared with the actual fluctuations in Eq.~\eqref{delTdecomp}, 
the modulation term in Eq.~\eqref{delTinfer} has taken on a peculiar 
frequency dependence, represented by $\fnu$.  This has arisen due to the 
coupling between the fluctuations and the dipole, $T_0\hatn\cdot\vec{\beta}$, 
which leads to a second-order term in the expansion of Eq.~\eqref{Inuexpn}.  
Intuitively, the CMB temperature varies from one side of the sky to 
the other at the 3\,mK level.  Therefore so does the calibration factor 
$dI_\nu/dT$, as represented by the second derivative $d^2I_\nu/dT^2$.  
We note that such a frequency-dependent modulation is not uniquely a velocity
effect, but would have arisen in the presence of any sufficiently large
temperature fluctuation.
Of course if we measured $T(\hatn)$ directly (for example by measuring 
$I_\nu(\nu, \hatn)$ at a large number of frequencies), we would measure 
the true fluctuations, Eq.~\eqref{delTdecomp}, i.e., we would have a boost 
factor of $\fnu = 1$.  However, this is not what happens in practice, and
hence the velocity-driven modulation has a spectrum which mixes in a 
$d^2I_\nu/dT^2$ dependence.

\section{Methodology}

\begin{figure}[!t]
\centerline{\includegraphics[width=0.9\columnwidth]{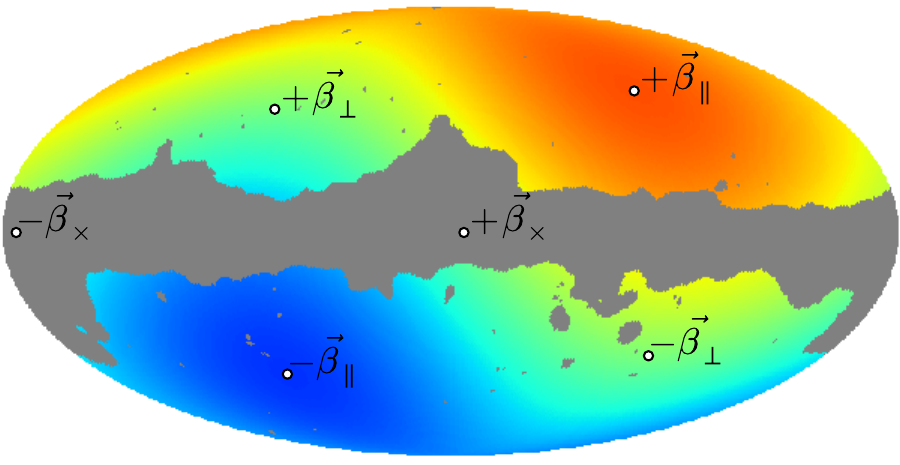}}

\caption{
Specific choice for the decomposition of the dipole vector
$\vec{\beta}$ in Galactic coordinates. 
The CMB dipole direction $(l,b) =(263\pdeg99, 48\pdeg26)$ is given as
$\vec{\beta}_{\parallel}$,
while two directions orthogonal to it (and each other) are denoted as
$\vec{\beta}_{\perpa}$ and $\vec{\beta}_{\perpb}$.  The vector 
$\vec{\beta}_{\times}$ lies within the Galactic plane.
}
\label{fig:dipole_components}
\end{figure}

The statistical properties of the aberration-induced stretching and compression of the CMB anisotropies
are manifest in ``statistically anisotropic''
contributions to the covariance matrix of the CMB, which we can use to
reconstruct the  velocity vector \citep{Burles:2006xf,Kosowsky:2010jm,Amendola:2010ty}.
To discuss these it will be useful to introduce the harmonic transform of the
peculiar velocity vector, given by
\begin{align}
\beta_{LM} 
&= \int d\hatn Y_{LM}^*(\hatn) \vec{\beta} \cdot \hatn.
\end{align}
Here $\beta_{LM}$ is only non-zero for dipole modes (with $L=1$).
Although most of our equations will be written in terms of $\beta_{LM}$,
for simplicity of interpretation we will present results in a specific choice
of basis for the three dipole modes of orthonormal unit vectors, labelled
$\vec{\beta}_{\parallel}$ (along the expected velocity direction), 
$\vec{\beta}_{\perpb}$ (perpendicular to $\vec{\beta}_{\parallel}$ and parallel to 
the Galactic plane, near its centre), and the remaining perpendicular mode 
$\vec{\beta}_{\perpa}$.
The directions associated with these modes are plotted
in Fig.~\ref{fig:dipole_components}. 

In the statistics of the CMB fluctuations, peculiar velocity effects manifest
themselves as a set of off-diagonal contributions to the CMB covariance matrix:
\begin{equation}
\langle T_{\ell_1 m_2} T_{\ell_2 m_2} \rangle_{\mathsc{cmb}} = 
\sum_{LM} 
(-1)^M
\threej{\ell_1}{\ell_2}{L}{m_1}{m_2}{-M} 
W_{\ell_1 \ell_2 L}^{\betanu}\,\beta_{LM},
\end{equation}
where the weight function $W^{\betanu}$ is composed of two parts, related
to the \aberration\ and \modulation\ effects, respectively,
\be
W^{\betanu}_{\ell_1 \ell_2 L} = W^{\phi}_{\ell_1 \ell_2 L} + \fnu
 W^{\tau}_{\ell_1 \ell_2 L},
 \label{eqn:wbetanu}
\ee
and the term in large parentheses is a Wigner 3-$j$ symbol.
It should be understood that in all of the expressions in this 
section, we take $L=1$ for our calculations.  We have written the expressions 
in more general form to allow easier connection to more general estimators 
in the literature.
The \aberration\ term, for example, is identical to that found when
considering gravitational lensing of the CMB by large-scale
structure \citep{Lewis:2006fu,planck2013-p12},
\begin{multline}
W_{\ell_1 \ell_2 L}^{\phi} 
=  \left( \frac{1 + (-1)^{\ell_1 + \ell_2 + L}}{2} \right)
\sqrt{\frac{(2\ell_1+1)(2\ell_2+1)(2L+1)}{4\pi}} \\
\times  \sqrt{ L(L+1) \ell_1 (\ell_1+1) } 
 C_{\ell_1}^{\rm TT} \threej{\ell_1}{\ell_2}{L}{1}{0}{-1}
 + (\ell_1 \leftrightarrow \ell_2),
\label{eqn:qe_weight_lensing}
\end{multline}
while the \modulation\ term is identical to that produced by an inhomogeneous
optical depth to the last scattering surface \citep{Dvorkin:2008tf},
\be
W_{\ell_1 \ell_2 L}^{\tau}
= \sqrt{\frac{(2\ell_1+1)(2\ell_2+1)(2L+1)}{4\pi}}
\threej{\ell_1}{\ell_2}{L}{0}{0}{0} \left( C_{\ell_1}^{\rm TT}
 + C_{\ell_2}^{\rm TT} \right).
\ee
Note that one might, in principle, be concerned about order $\beta^2$
corrections to the covariance matrix, particularly at high $\ell$ 
\citep[see, e.g.,][]{Notari:2011sb}.  However, these are small, provided
that the spectra are relatively smooth.  Although the order
$\beta$ ($\simeq 10^{-3})$ deflections give large changes to the $a_{\ell m}$s
for $\ell>10^3$, the changes to the overall covariance are small 
\citep{chluba:2011zh}, since the deflection effect is coherent over very
large scales.

The basic effect of these
boosting-induced correlations is to couple $\ell$ modes with $\ell \pm 1$ 
modes.  They therefore share this property with pure dipolar amplitude 
modulations studied in the context of primordial statistical anisotropy 
\citep[see, e.g,][]{prunet2005}, as well as with dipolar modulations in more 
general physical parameters \citep{moss2011}.  However, these other cases 
do not share the frequency dependence of the boosting modulation effect, 
since they are not accompanied by a temperature dipole.

We measure the peculiar velocity dipole using quadratic estimators,
essentially summing over the covariance matrix of the observed CMB
fluctuations, with weights designed to optimally extract $\vec{\beta}$.
A general quadratic estimator 
$\hat{x\,}_{LM}$ for $\vec{\beta}_{LM}$
is given by 
\citep{Hanson:2009gu}
\begin{multline}
\hat{x\,}_{LM}[\bar{T}] =
\frac{1 }{2} N_L^{x \betanu} 
\sum_{\ell_1= \ell_{\rm min}}^{\ell_{\rm max}}
 \sum_{\ell_2= \ell_{\rm min}}^{\ell_{\rm max}}  \sum_{m_1, m_2} 
 (-1)^M
\threej{\ell_1}{\ell_2}{L}{m_1}{m_2}{-M} W_{\ell_1 \ell_2 L}^{{x\,}}
  \\ \times
\left( 
\bar{T}^{}_{\ell_1 m_1} \bar{T}^{}_{\ell_2 m_2} -
\langle \bar{T}^{}_{\ell_1 m_1} \bar{T}^{}_{\ell_2 m_2}  \rangle
\right),
\label{eqn:qe_block}
\end{multline}
where $\bar{T}_{\ell m}$ are a set of inverse-variance filtered temperature
multipoles, $W^{x}_{\ell_1 \ell_2 L}$ is a weight function and
$N_L^{x \betanu}$ is a normalization.
To study the total boosting effect we use Eq.~\eqref{eqn:wbetanu} 
for the weight function, but we will also use weight functions designed to 
extract specifically the aberration and modulation components of the effect.
The ensemble average term $\langle \rangle$ is taken over signal+noise
realizations of the CMB in the absence of velocity effects. 
It corrects for the statistical anisotropy induced by effects like beam
asymmetry, masking, and noise inhomogeneity.  We evaluate this term using
Monte Carlo simulations of the data, as discussed in
Sect.~\ref{sec:simulations}.

We use three different quadratic estimators to measure the effects of boosting.
The first, $\hat{\beta\,}$, simply adopts the weight function
$W^{\betanu}_{\ell_1 \ell_2 L}$, and provides a minimum-variance estimator
of the total peculiar velocity effect.
The two additional estimators, $\hat{\phi\,}$ and $\hat{\tau\,}$, isolate the
\aberration\ and \modulation\ aspects of the peculiar velocity effect,
respectively.
This can be useful, as they are qualitatively quite distinct effects, and
suffer from different potential contaminants.  The modulation effect, for
example, is degenerate with a dipolar pattern of calibration errors on the sky,
while the aberration effect is indistinguishable from a dipolar pattern of
pointing errors.

There is a subtlety in the construction of these estimators, due to the fact
that the covariances, described by $W^{\phi}$ and $W^{\tau}$, are not
orthogonal.  To truly isolate the \aberration\ and \modulation\ effects, we
form orthogonalized weight matrices as
\begin{align}
W^{\hat{\phi\,}}_{\ell_1 \ell_2 L} &= W^{\phi}_{\ell_1 \ell_2 L} -
 W^{\tau}_{\ell_1 \ell_2 L} \frac{{\cal R}_L^{\phi \tau}}
 {{\cal R}_L^{\tau \tau}} \quad {\rm and} \label{eqn:phibhwt}\\
W^{\hat{\tau\,}}_{\ell_1 \ell_2 L} &= W^{\tau}_{\ell_1 \ell_2 L} -
 W^{\phi}_{\ell_1 \ell_2 L} \frac{{\cal R}_L^{\tau \phi}} {{\cal R}_L^{\phi \phi}},
\label{eqn:taubhwt}
\end{align}
where the response function ${\cal R}$ is given by
\be
{\cal R}_L^{x z} =  \frac{1}{(2L+1)} \sum_{\ell_1 =
 \ell_{\rm min}}^{\ell_{\rm max}} \sum_{\ell_2 =
 \ell_{\rm min}}^{\ell_{\rm max}}
 \frac{1}{2} W_{\ell_1 \ell_2 L}^{x} W_{\ell_1 \ell_2 L}^{z}
 F_{\ell_1}^{} F_{\ell_2}^{},
\ee
with $x,z = \betanu,\phi,\tau$.  
The construction of these estimators is analogous to the construction of
``bias-hardened'' estimators for CMB lensing \citep{Namikawa:2012pe}.
The spectra $F_{\ell}$ are diagonal approximations to the inverse variance
filter, which takes the sky map $T \rightarrow \bar{T}$.
We use the same inverse variance filter as that used for the baseline results
in \cite{planck2013-p12}, and the approximate filter functions are also
specified there.
Note that our $\hat{\phi\,}$ estimator is slightly different from that
used in \cite{planck2013-p12}, due to the fact that we have
orthogonalized it with respect to $\tau$. 

The normalization $N_L^{x \betanu}$ can be approximated analytically as
\be
N_L^{x \betanu} \simeq \left[  {\cal R}_L^{x \betanu} \right]^{-1}.
\label{eqn:nlxbnu}
\ee
This approximation does not account for masking.  On a masked sky, with this
normalization, we expect to find that
$\left< \hat{x\,}_{LM} \right> = f_{LM,\,{\rm sky}}\, \beta_{LM}$, where 
\be
f_{LM,\,{\rm sky}} = \int d\hatn Y_{LM}^*(\hatn) M(\hatn)
 \vec{\beta}_{\parallel} \cdot \hatn.
\label{eqn:fsky}
\ee
Here $M(\hatn)$ is the sky mask used in our analysis.
For the fiducial sky mask we use (plotted in Fig.~\ref{fig:dipole_components}, 
and which leaves approximately $70\%$ of the sky unmasked), 
taking the dot product of $f_{1M,\,{\rm sky}}$ with
our three basis vectors we find that
$f_{\parallel,\ {\rm sky}} = 0.82$, $f_{\perp,\ {\rm sky}} = 0.17$, and 
$f_{\times,\ {\rm sky}} = -0.04$.  The large effective sky fraction for the 
$\vec{\beta}_{\parallel}$ direction reflects the fact that the peaks of the 
expected velocity dipole are untouched by the mask, while the small values of 
$f_{\rm sky}$ for the other components reflects that fact that the masking 
procedure does not leak a large amount of the dipole signal in the 
$\vec{\beta}_{\parallel}$ direction into other modes.

\section{Data and simulations}
\label{sec:dataandsims}
\label{sec:data}
\label{sec:simulations}

Given the frequency-dependent nature of the velocity effects we are searching
for (at least for the $\tau$ component), we will focus for the most part on
estimates of $\beta$ obtained from individual frequency maps, although in
Sect.~\ref{page:fgnds} we will also discuss the analysis of component-separated
maps obtained from combinations of the entire \Planck\ frequency range.
Our analysis procedure is essentially identical to that of
\cite{planck2013-p12}, and so we only provide a brief review of it here.
We use the 143 and 217\,GHz \Planck\ maps, which contain the majority
of the available CMB signal probed by \Planck\ at the high multipoles required
to observe the velocity effects.
The 143\,GHz map has a noise level that is reasonably well approximated by
$45\,\mu{\rm K}\,{\rm arcmin}$ white noise, while the 217\,GHz map has
approximately twice as much noise power, with a level of
$60\,\mu{\rm K}\,{\rm arcmin}$.
The beam at 143\,GHz is approximately $7^\prime$ FWHM, while the 217\,GHz
beam is $5^\prime$ FWHM.
This increased angular resolution, as well as the larger size of the
$\tau$-type velocity signal at higher frequency, makes 217\,GHz slightly more
powerful than 143\,GHz for detecting velocity effects (the HFI 100\,GHz
and LFI 70\GHz\ channels would offer very little additional constraining power).
At these noise levels, for $70\%$ sky coverage we Fisher-forecast a $20\%$
measurement of the component $\beta_{\parallel}$ at 217\,GHz
(or, alternatively, a $5\,\sigma$
detection) or a $25\%$ measurement of $\beta_{\parallel}$ at 143\,GHz,
consistent with the estimates of \cite{Kosowsky:2010jm} and \cite{Amendola:2010ty}.
As we will see, our actual statistical error bars determined from
simulations agree well with these expectations. 

The \Planck\ maps are generated at
{\tt HEALPix} \citep{gorski2005}\footnote{\url{http://healpix.jpl.nasa.gov}}
$N_{\rm side} = 2048$.
In the process of mapmaking, time-domain observations are binned into pixels.
This effectively generates a pointing error, given by the distance between
the pixel centre to which each observation is assigned and the true pointing
direction at that time.
The pixels at $N_{\rm side}=2048$ have a typical dimension of $1\parcm7$. 
As this is comparable to the size of the aberration effect we are looking for, this is a potential source of concern.
However, as discussed in \cite{planck2013-p12}, the
beam-convolved CMB is sufficiently smooth on these scales that it is well
approximated as a gradient across each pixel, and the errors accordingly
average down with the distribution of hits in each pixel.
For the frequency maps that we use, the rms pixelization error is on the 
order of $0\parcm1$, and not coherent over the large dipole scales which
we are interested in, and so we neglect pixelization effects in our measurement.

We will use several data combinations to measure $\vec{\beta}$.
The quadratic estimator of Eq.~\eqref{eqn:qe_block} has two input ``legs,''
i.e., the $\ell_1 m_1$ and $\ell_2 m_2$ terms.  
Starting from the 143 and 217\,GHz maps, there are three distinct ways
we may source these legs:
(1) both legs use either the individually filtered 143\,GHz or 217\,GHz maps,
which we refer to as
$143\times 143$ and $217 \times 217$, respectively;
(2) we can use 143\,GHz for one leg, and 217\,GHz for the other, 
referred to as $143 \times 217$; and
(3) we can combine both 143 and 217\,GHz data in our inverse-variance filtering
procedure into a single minimum-variance map, which is then fed into both
legs of the quadratic estimator. We refer to this final combination
schematically as ``143+217.''  Combinations (2) and (3) mix 143 and 217\,GHz
data.  When constructing the weight function of Eq.~\eqref{eqn:wbetanu} for
these combinations we use an effective $\fnu=2.5$. 
Note that this effective $\fnu$ is only used to determine the weight function
of the quadratic estimator; errors in the approximation will make our
estimator suboptimal, but will not bias our results.
To construct the $\bar{T}$, which are the inputs for these quadratic
estimators, we use the filtering described in Appendix~A of
\cite{planck2013-p12}, which optimally accounts for the Galactic and point
source masking (although not for the inhomogeneity of the instrumental noise
level).
This filter inverse-variance weights the CMB fluctuations, and also projects
out the 857\,GHz \Planck\ map as a dust template.

To characterize our estimator and to compute the mean-field term of
Eq.~\eqref{eqn:qe_block}, we use a large set of Monte Carlo simulations.
These are generated following the same procedure as those described in
\cite{planck2013-p12};
they incorporate the asymmetry of the instrumental beam, gravitational lensing
effects, and realistic noise realizations from the FFP6 simulation set
described in \cite{planck2013-p01} and \cite{planck2013-p28}.
There is one missing aspect of these simulations which we discuss briefly here:
due to an oversight in their preparation,
the gravitational lensing component of our simulations only included
lensing power for lensing modes on scales $L \ge 2$, which leads to
a slight underestimation of our simulation-based error bars for the
$\phi$ component of the velocity estimator.
The lensing dipole power in the fiducial $\Lambda$CDM model is 
$C_1^{\phi\phi} \simeq 6\times 10^{-8}$, which represents an additional
source of noise for each mode of $\vec\beta$, given by 
$\sigma_{\phi, {\rm lens}} = 1.2 \times 10^{-4}$, or about one tenth
the size of the expected signal. 
The $\phi$ part of the estimator contributes approximately $46\%$ of the
total $\beta$ estimator weight at 143\,GHz, and $35\%$ at 217\,GHz.
Our measurement errors \textit{without} this lensing noise on an individual
mode of $\vec\beta$ are $\sigma_{\beta} \simeq 2.5 \times 10^{-4}$, while with
lensing noise included we would expect this to increase to 
$\sqrt{ \sigma^2_{\beta} + (4/10)^2 \sigma_{\phi, {\rm lens}}^2 }
 = 2.54 \times 10^{-4}$.
This is small enough that we have neglected it for these
results (rather than include it by hand).

We generate simulations both with and without peculiar velocity effects, to
determine the normalization of our estimator, which, as we will see, is
reasonably consistent with the analytical expectation discussed around
Eq.~\eqref{eqn:fsky}.
All of our main results with frequency maps use 1000 simulations to determine
the estimator mean field and variance, while the component separation
tests in Sect.~\ref{page:fgnds} use 300 simulations.

\section{Results}
\begin{figure*}[!ht]
\centerline{\includegraphics[width=\textwidth]{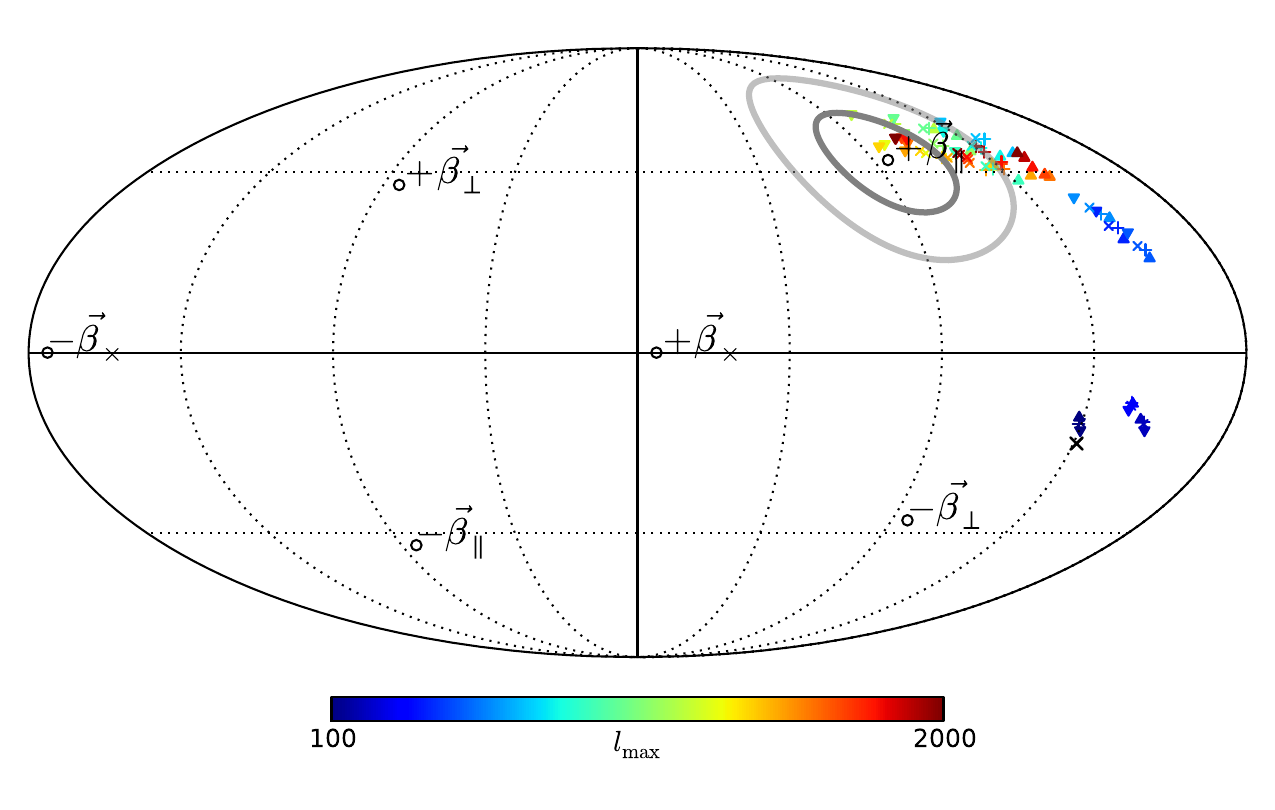}}
\caption{
Measured dipole direction $\hat{\vec{\beta\,}}$ in Galactic coordinates as a
function of the maximum temperature multipole used in the analysis,
$\ell_{\rm max}$.
We plot the results for the four data combinations discussed in
Sect.~\ref{sec:dataandsims}: 
$143 \times 143$ ($\blacktriangledown$ symbol);
$217 \times 217$ ($\blacktriangle$ symbol);
$143 \times 217$ ($\times$ symbol); and $143+217$ ($+$ symbol).
The CMB dipole direction $\vec{\beta}_{\parallel}$ has been highlighted with
$14^\circ$ and $26^\circ$ radius circles, which correspond roughly to our
expected uncertainty on the dipole direction.
The black cross in the lower hemisphere is the modulation dipole anomaly
direction found for {\it WMAP\/} at $\ell_{\rm max}=64$ in
\cite{Hoftuft:2009rq}, and which is discussed further in \cite{planck2013-p09}.
Note that all four estimators are significantly correlated with one another,
even the $143 \times 143$ and $217 \times 217$ results, which are based on
maps with independent noise realizations.  This is because a significant
portion of the dipole measurement uncertainty is from sample variance of
the CMB fluctuations, which is common between channels.
}
\label{fig:pecvec_mollview}
\end{figure*}

We present our results visually in Fig.~\ref{fig:pecvec_mollview}, where we
plot the total measured dipole direction $\hat{\vec{\beta\,}}$ as a function
of the
maximum temperature multipole $\ell_{\rm max}$ used as input to our quadratic
estimators.
We can see that all four of our 143/217\,GHz based estimators converge toward
the expected dipole direction at high $\ell_{\rm max}$.
At $\ell_{\rm max} < 100$, we recover the significantly preferred direction
of \cite{Hoftuft:2009rq}, which is identified when searching for a dipolar
modulation of the CMB fluctuations \citep[see][]{planck2013-p09}.
The $\tau$ component of the velocity effect is degenerate with such a
modulation (at least at fixed frequency), and $\phi$ gets little weight 
from $\ell < 100$, so this is an expected result.
The significance of this preferred direction varies as a function of
smoothing scale 
\citep{Hanson:2009gu,Bennett:2010jb,planck2013-p09}.
To minimize possible contamination of our results by this potential anomaly,
from here onward we restrict the temperature multipoles used in our $\beta$
estimation to $\ell_{\rm min} = 500$. This cut removes only about 10\% of
the total number of modes measured by $\Planck$, and so does not
significantly increase our error bars.  Note also that we have 
verified that our error bars do not shrink significantly for 
$\ell_{\rm max} > 2000$, since almost all of the modes measured by Planck 
are at $\ell < 2000$.

There is a clear tendency in Fig.~\ref{fig:pecvec_mollview} for the measured
velocity to point toward the expected direction $\vec{\beta}_{\parallel}$.
At \Planck\ noise levels, we expect a $1\,\sigma$ uncertainty on
each component of $\vec{\beta}$ of better than $25\%$.
A $25\%$ uncertainty corresponds to an $\arctan( 1/4 ) = 14^\circ$
constraint on the direction of $\vec{\beta}$. 
We plot this contour, as well as the corresponding $2\,\sigma$
contour $\arctan(2/4) = 26^\circ$. 
It is apparent that the measured velocity directions are in reasonable
agreement with the CMB dipole.

\begin{figure*}[!htpb]
\vspace{0.0in}
\centerline{\includegraphics[width=\textwidth]{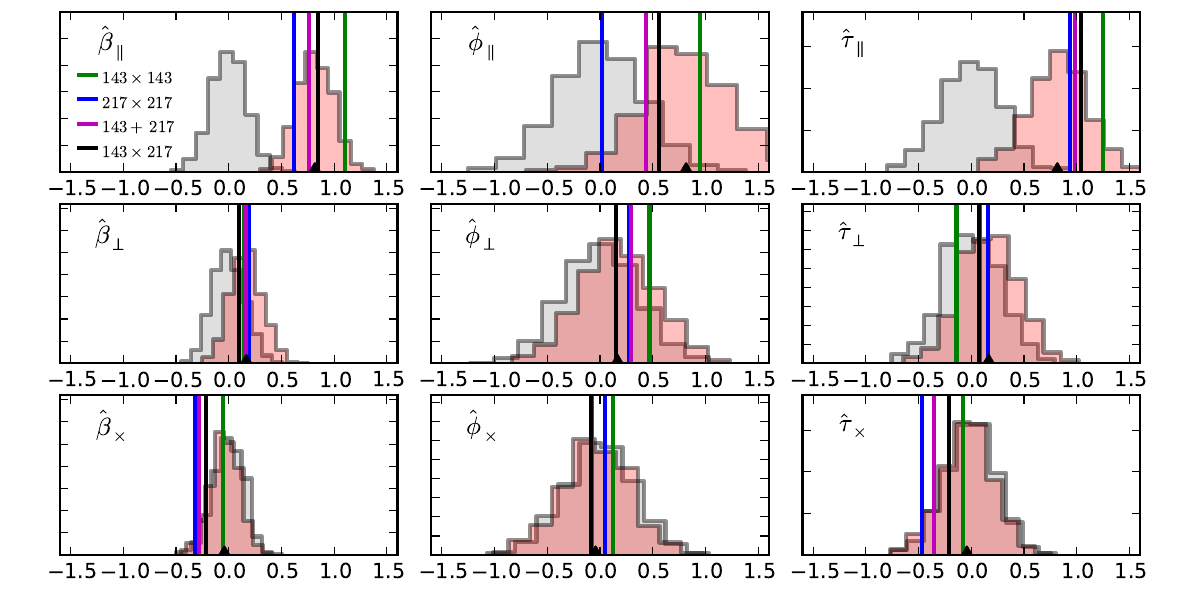}}
\caption{
Measurements of $\vec{\beta}$ using combinations of the 143 and 217\,GHz \Planck\ maps, 
normalized using Eq.~\eqref{eqn:nlxbnu} and then divided by
the fiducial amplitude of $\beta = 1.23 \times 10^{-3}$.
These estimates use $\ell_{\rm min}=500$ and $\ell_{\rm max}=2000$.
In addition to the total minimum variance estimate $\hat{\beta\,}$,
the measurement is also broken down into its \aberration-type part, 
$\hat{\phi\,}$, and \modulation-type part, $\hat{\tau\,}$.
Vertical lines give the \Planck\ measurement for the four estimates
described in the text.
Grey histograms give the distribution of estimates for 
simulations of the $143\times217$ estimator, which do {\it not\/} contain 
peculiar velocity effects (the other estimators are very similar).
The red histograms give the distribution for simulations which {\it do\/}
contain peculiar velocity effects, simulated with the fiducial direction
(along $\vec{\beta}_\parallel$) and amplitude.
Black triangles on the $x$-axis indicate the relevant component of
$f_{{\rm sky}}$ given by Eq.~\eqref{eqn:fsky}, which agrees well
with the peak of the velocity simulations.
}
\label{fig:pecvec_grid_freq}

\vspace{0.25in}
\centerline{\includegraphics[width=\textwidth]{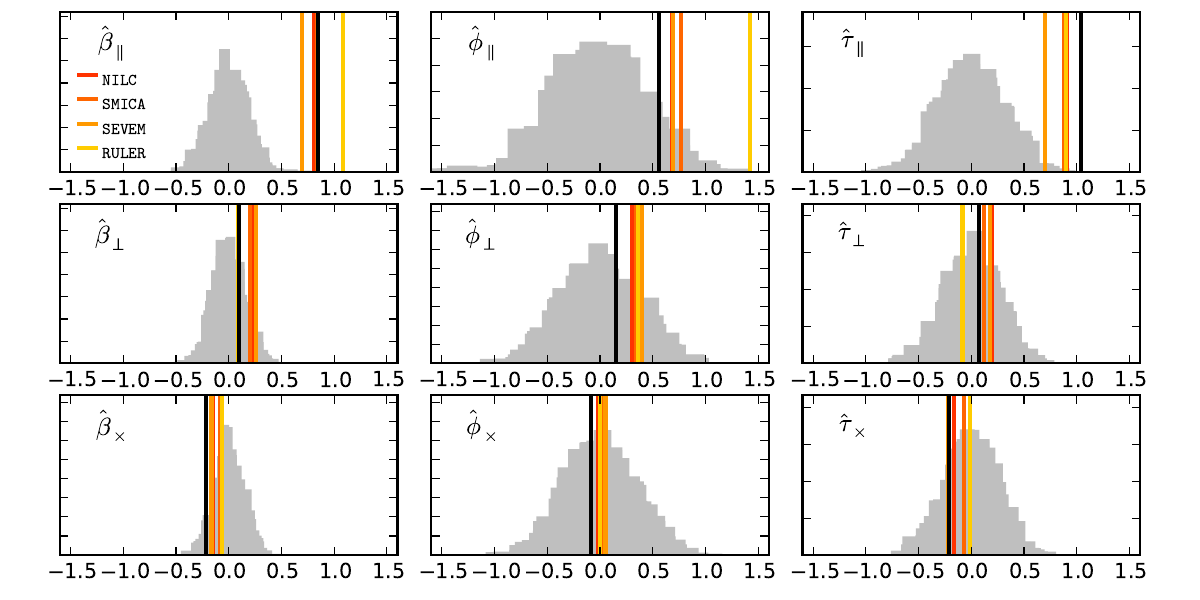}}
\caption{
Plot of velocity amplitude estimates, similar to
Fig.~\ref{fig:pecvec_grid_freq}, but using an array of
component-separated maps, rather than specific combinations of frequency maps.
The production and characterization of these component-separated maps is
presented in \cite{planck2013-p06}.
Histograms of simulation results \textit{without} velocity effects are
overplotted in grey for each method; they are all very similar.
Vertical coloured bars correspond to the maps indicated in the legend,
using the combination
of our fiducial galaxy mask (which removes approximately $30\%$ of the sky), as
well as the specific mask produced for each component separation method.
We see significant departures from the null-hypothesis simulations only in 
the $\vec{\beta}_{\parallel}$ direction, as expected.
Vertical black lines show the $143 \times 217$ measurement of
Fig.~\ref{fig:pecvec_grid_freq}.
Note the discussion about the subtleties in the normalization of
these estimates in Sect.~\ref{page:fgnds}.
}
\label{fig:pecvec_grid_csep}
\end{figure*}

We now proceed to break the measurement of Fig.~\ref{fig:pecvec_mollview}
into its constituent parts for $\ell_{\rm max} = 2000$ (and truncating
now at $\ell_{\rm min} = 500$).  In Fig.~\ref{fig:pecvec_grid_freq}
we plot our quadratic estimates of the three components of $\vec{\beta}$,
as well as the decomposition into \aberration\ and \modulation\ components, 
for each of our four frequency combinations.
The vertical lines in Fig.~\ref{fig:pecvec_grid_freq}
give the amplitude estimates for each component measured from the data,
while the coloured and grey histograms give the distribution of these
quantities for the $143\times217$
estimator, for simulations with and without velocity effects, respectively (the 
other estimators are similar).  
As expected, the velocity effects show up primarily in $\beta_{\parallel}$;
there is little leakage into other components with our sky mask.
For all four estimators, we see that the presence of velocity
along $\vec{\beta}_{\parallel}$ is strongly preferred over the null hypothesis. 
At 143\,GHz this signal comes from both $\hat{\phi\,}_{\parallel}$ and
$\hat{\tau\,}_{\parallel}$.
At 217\,GHz it comes primarily from $\hat{\tau\,}_{\parallel}$.
Additionally, there is a somewhat unexpected signal at $217\,$GHz in the
$\vec{\beta}_{\times}$ direction, again driven by the $\tau$ component.
Given the apparent frequency dependence, foreground contamination seems a
possible candidate for this anomalous signal.
We will discuss this possibility further in the next section.

In Table~\ref{table:bparsignifs} we present $\chi^2$ values for the
$\vec{\beta}$ measurements of Fig.~\ref{fig:pecvec_grid_freq}
under both the null hypothesis of no velocity effects, and
assuming the expected velocity direction and amplitude.
We can see that all of our measurements are in significant disagreement
with the ``no velocity'' hypothesis. The probability-to-exceed (PTE) values for
the ``with velocity'' case are much more reasonable. 
Under the velocity hypothesis, $217\times217$ has the lowest PTE, of
$11\%$, driven by the large $\hat{\beta\,}_{\times}$. 

In Table~\ref{table:bparamps} we focus on our measurements of the velocity
amplitude along the expected direction $\vec{\beta}_{\parallel}$, as well as
performing null tests among our collection of estimates.
For this table, we have normalized the estimators, such that
the average of $\hat{\beta\,}_{\parallel}$ on boosted simulations is equal to
the input value of $369\kms$.
For all four of our estimators, we find that this normalization factor is
within $0.5\%$ of that given by $N^{x \beta_{\nu}} f_{\parallel, {\rm sky}}$,
as is already apparent from the triangles
along the horizontal axis of Fig.~\ref{fig:pecvec_grid_freq}. 
We can see here, as expected, that our estimators have a statistical
uncertainty on $\beta_{\parallel}$ of between $20\%$ and $25\%$.  However,
several of our null tests, obtained by taking the differences of pairs of
$\beta_{\parallel}$ estimates, fail at the level of 2--$3\,\sigma$.
We take the $143\times217$\,GHz estimator as our fiducial measurement;
because it involves the cross-correlation of two maps with independent noise
realizations it should be robust to noise modelling.
Null tests against the individual 143 and 217\,GHz estimates are in tension at
a level of $2\,\sigma$ for this estimator.
We take this tension as a measure of the systematic differences between
these two channels, and conservatively choose the largest discrepancy with the 
$143\times217$\,GHz estimate, namely $0.31$, as our systematic error.  We 
therefore report a measurement of 
$\hat{v\,}_{\parallel} = 384\kms \pm 78\kms\ ({\rm stat.}) \pm 115\kms \
({\rm syst.})$, a significant confirmation of the expected velocity amplitude.
\begin{table}[!tbh]
\begingroup
\newcommand{\mcg}{}
\newcommand{\scg}{\white} 
\newcommand{\dcg}{}
\newdimen\tblskip \tblskip=5pt
\caption{
Significance measures for the $\vec{\beta}$ estimates of
Fig.~\ref{fig:pecvec_grid_freq}.
We form a $\chi^2$ for the three measured modes of $\vec{\beta}$,
using the mean and covariance matrix measured for simulations either
with or without velocity effects.  The covariance matrices are very similar
in both cases, and so the difference between these two cases is
only the mean-field subtraction.
The ``PTE'' columns give the corresponding probability-to-exceed values
for a $\chi^2$ distribution with three degrees of freedom.
The measured $\chi^2$ values are extremely unlikely under the ``no velocity''
hypothesis, but very compatible for the ``with velocity'' case.
}
\label{table:bparsignifs}
\nointerlineskip
\vskip -3mm
\footnotesize
\setbox\tablebox=\vbox{
 \newdimen\digitwidth
 \setbox0=\hbox{\rm 0}
  \digitwidth=\wd0
  \catcode`*=\active
  \def*{\kern\digitwidth}
  \newdimen\signwidth
  \setbox0=\hbox{+}
  \signwidth=\wd0
  \catcode`!=\active
  \def!{\kern\signwidth}
\halign{\hbox to 2.0cm{$#$\leaderfil}\tabskip=0.5em&
  \hfil$#$\hfil\tabskip=1.5em&
  \hfil$#$\hfil\tabskip=2.5em&
  \hfil$#$\hfil\tabskip=1.5em&
  \hfil$#$\hfil\tabskip=0pt\cr
\noalign{\doubleline}
\omit&\multispan4\hfil $\hat{\beta\,}$ significance\hfil\cr
\noalign{\vskip -3pt}
\omit&\multispan4\hrulefill\cr
\noalign{\vskip 1pt}
\omit&\multispan2\hfil No velocity\hfil&\multispan2\hfil With velocity\hfil\cr
\noalign{\vskip 4pt\hrule\vskip 5pt}
\omit&\chi^2 & {\rm PTE}\ [\%] & \chi^2  & {\rm PTE}\ [\%]\cr
\noalign{\vskip 4pt\hrule\vskip 5pt}
143\times 143&27.1 & 0.0005 &  1.9 &  58.87\cr
217\times 217&20.7 & 0.0123 &  6.0 &   11.18\cr
143+217&25.1 &  0.0015 &  3.3 &   35.44\cr
143\times 217&27.6 & 0.0005 &  1.8 &  62.29\cr
\noalign{\vskip 4pt\hrule\vskip 3pt}}}
\endPlancktable
\endgroup
\end{table}
\begin{table}[!tbh]
\begingroup
\newcommand{\mcg}{}
\newcommand{\scg}{\white} 
\newcommand{\dcg}{}
\newdimen\tblskip \tblskip=5pt
\caption{Measured velocity amplitudes along the $\vec{\beta}_{\parallel}$
direction, in units of $369\kms$, using combinations of the 143 and 217\,GHz
data, as discussed in Sect.~\ref{sec:dataandsims}.  
The diagonal shows the results for the indicated reconstruction.  
Below the diagonal, the numbers given are for the difference of the two
results, and the uncertainty accounts for the correlation between each pair of
measurements.  This lower triangle is a null test.  
Several of these null tests fail between 143 and 217\,GHz, although both
channels provide evidence for $\beta_{\parallel}$, which is consistent with
the expected amplitude and discrepant with zero at 4--$5\,\sigma$.
}
\label{table:bparamps}
\nointerlineskip
\vskip -3mm
\footnotesize
\setbox\tablebox=\vbox{
 \newdimen\digitwidth
 \setbox0=\hbox{\rm 0}
  \digitwidth=\wd0
  \catcode`*=\active
  \def*{\kern\digitwidth}
  \newdimen\signwidth
  \setbox0=\hbox{+}
  \signwidth=\wd0
  \catcode`!=\active
  \def!{\kern\signwidth}
\halign{\hbox to 2.0cm{$#$\leaderfil}\tabskip=0.5em&
  \hfil$#$\hfil\tabskip=1.0em&
  \hfil$#$\hfil&
  \hfil$#$\hfil&
  \hfil$#$\hfil\tabskip=0pt\cr
\noalign{\doubleline}
\omit&\multispan4\hfil $\hat{\beta\,}_{\parallel}$ Amplitude\hfil\cr
\noalign{\vskip -3pt}
\omit&\multispan4\hrulefill\cr
\noalign{\vskip 1pt}
\omit&143\times143&217\times217&143+217&143\times217\cr
\noalign{\vskip 4pt\hrule\vskip 4pt}
143\times 143&\dcg1.35 \pm 0.26&\scg2.10 \pm 0.41&\scg2.27 \pm 0.44&\scg2.39 \pm 0.46\cr
217\times 217&\mcg0.60 \pm 0.21&\dcg0.75 \pm 0.19&\scg1.67 \pm 0.38&\scg1.79 \pm 0.38\cr
143+217&\mcg0.43 \pm 0.15&\mcg\llap{$-$}0.17 \pm 0.09&\dcg0.92 \pm 0.20&\scg1.96 \pm 0.40\cr
143\times 217&\mcg0.31 \pm 0.13&\mcg\llap{$-$}0.29 \pm 0.12&\mcg\llap{$-$}0.12 \pm 0.07&\dcg1.04 \pm 0.21\cr
\noalign{\vskip 4pt\hrule\vskip 3pt}}}
\endPlancktable
\endgroup
\end{table}

\section{Potential contaminants}

There are several potential sources of contamination for our estimates above which we discuss briefly here,
although we have not attempted an exhaustive study of potential contaminants for our estimator.

\textit{Galactic Foregrounds:}
\label{page:fgnds}
Given the simplicity of the foreground correction we have used (consisting
only of masking the sky and projecting out the \Planck\ 857\,GHz map as a crude dust
template), foreground contamination is a clear source of concern.
The frequency dependence of the large $\beta_{\times}$ signal seen at
217\,GHz, but not at 143\,GHz, seems potentially indicative of
foreground contamination, as the Galactic dust power is approximately
10 times larger at 217 than at 143\,GHz.
To test the possible magnitude of residual foregrounds, we apply our velocity
estimators to the four component-separated CMB maps of \cite{planck2013-p06},
i.e., {\tt NILC}, {\tt SMICA}, {\tt SEVEM}, and {\tt COMMANDER-RULER}. 
Each of these methods combines the full set of nine \Planck\ frequency maps from 30 to 857\,GHz to obtain a best-estimate CMB map.
To characterize the scatter and mean field of each method's map we use the
set of common simulations which each method has been applied to.
These simulations include the effect of the aberration part of the 
velocity dipole, although not the frequency-dependent modulation part.
For this reason, it is difficult to accurately assess the normalization of
our estimators when applied to these maps, particularly as they can mix
143 and 217\,GHz as a function of scale, and the modulation part is
frequency dependent.  We can, however, study them at a qualitative level.
The results of this analysis are shown in Fig.~\ref{fig:pecvec_grid_csep}.
To construct our $\vec{\beta}$ estimator for the component-separated maps
we have used $\fnu=2.5$, assuming that they contain roughly equal
contributions from 143 and 217\,GHz. 
Note that, because the simulations used to determine the mean fields of
the component-separated map included the \aberration\ part of the velocity
effect, it will be absorbed into the mean field if uncorrected.
Because the \aberration\ contribution is frequency independent
(so there are no issues with how the different
CMB channels are mixed), and given the good agreement between
our analytical normalization and that measured using simulations for the
frequency maps, when generating Fig.~\ref{fig:pecvec_grid_csep} we have
subtracted the expected velocity contribution from the mean field analytically.
We see generally good agreement with the $143\times217$ estimate on which we
have based our measurement of the previous section; there are no obvious
discrepancies with our
measurements in the $\vec{\beta}_{\parallel}$ direction, 
although there is a somewhat large scatter between methods for
$\hat{\phi\,}_{\parallel}$.
In the $\vec{\beta}_{\times}$ direction the component-separated map estimates 
agree well with the $143 \times 217$ estimator, and do not show the 
significant power seen for $217 \times 217$, suggesting that the large
power that we see there may indeed be foreground in origin.

\textit{Calibration errors:}
Position-dependent calibration errors in our sky maps are completely
degenerate with \modulation-type effects (at fixed frequency), and so are 
very worrisome as a potential systematic effect.
We note that the \Planck\ scan strategy strongly suppresses map calibration
errors with large-scale structure
\citep[such as a dipole, see][]{planck2013-p03f}. 
As the satellite spins, the detectors mounted in the focal plane inscribe
circles on the sky, with opening angles of between $83^\circ$ and
$85^\circ$ (for the 143 and 217\,GHz detectors we use).
For a time-dependent calibration error to project cleanly into a dipolar
structure on the sky, it would need to have a periodicity comparable to the
spin frequency of the satellite (1\,min$^{-1}$).
Slower fluctuations in the calibration should be strongly suppressed in
the maps.  There are calibration errors which are not suppressed by the
scan strategy, however.  For example, a nonlinear detector response could
couple directly to the large CMB dipole temperature. 
Ultimately, because position-dependent calibration errors are completely
degenerate with the $\tau$ component of the velocity effect, the only
handle which we have on them for this study is the consistency between
$\hat{\phi\,}$ and $\hat{\tau\,}$.  From another viewpoint, the consistency of
our measurement with the \textit{expected}
velocity \modulation\ provides an upper bound on dipolar calibration errors.

\textit{Pointing errors:}
In principle, errors in the pointing are perfectly degenerate with the
\aberration-type velocity effect in the observed CMB.  However
the \Planck\ pointing solution has an uncertainty of a few arcseconds
rms in both the co- and cross-scan directions \mbox{\citep{planck2013-p03}}.
The $3\arcmin$ rms aberrations induced by velocity effects are simply too
large to be contaminated by any reasonable pointing errors.

\section{Conclusions}
From Fig.~\ref{fig:pecvec_mollview} it is clear that small-scale CMB
fluctuations observed in the \Planck\ data provide evidence for velocity
effects in the expected direction.
This is put on more quantitative footing in Fig.~\ref{fig:pecvec_grid_freq}
and Table~\ref{table:bparamps}, where we see that all four of the
143 and 217\,GHz velocity estimators which we have considered show
evidence for velocity effects along $\vec{\beta}_{\parallel}$ 
at above the $4\,\sigma$ level.
Detailed comparison of 143 and 217\,GHz data shows some discrepancies, which
we have taken as part of a systematic error budget; however, tests with
component-separated maps shown in Fig.~\ref{fig:pecvec_grid_csep} provide a
strong indication that our 217\,GHz map has slight residual foreground
contamination.  The component-separated results are completely consistent
with the $143 \times 217$ estimator which we quote for our fiducial result.

Beyond our peculiar velocity's impact on the CMB, there have been many 
studies of related effects at other wavelengths \citep[e.g.,][]{Blake:2002,
Titov:2011,Gibelyou:2012}.  Closely connected are observational studies 
which examine the convergence of the clustering dipole 
\citep[e.g.,][]{Itoh:2010,bilicki2011}.  Indications of non-convergence might
be evidence for a super-Hubble isocurvature mode, which can generate a 
``tilt'' between matter and radiation \citep{turner1991}, leading to an 
extremely large-scale bulk flow.  Such a long-wavelength isocurvature 
mode could also contribute a significant ``intrinsic'' component to our 
observed temperature dipole \citep{langlois1996}.  However, a peculiar 
velocity dipole is {\it expected\/} at the level of $\sqrt{C_1} \sim 10^{-3}$
due to structure in standard $\Lambda$CDM \citep[see, e.g.,][]{Zibin:2008},
which suggests that an intrinsic component, if it exists at all,
is subdominant.  In addition, 
such a bulk flow has been significantly constrained by \Planck\ studies 
of the kinetic Sunyaev-Zeldovich effect \citep{planck2013-XIII}.  In 
this light, the observation of aberration at the expected level reported 
in this paper is fully consistent with the standard, adiabatic picture 
of the Universe.

The Copernican revolution taught us to see the Earth as orbiting a stationary 
Sun.  That picture was eventually refined to include Galactic and 
cosmological motions of the Solar System.  Because of the technical 
challenges, one may have thought it very unlikely to be able to measure (or 
perhaps even to {\em define}) the cosmological motion of the Solar
System \dots\ {\em and yet it moves.}

\begin{acknowledgements}
The development of \Planck\ has been supported by: ESA; CNES and
CNRS/INSU-IN2P3-INP (France); ASI, CNR, and INAF (Italy); NASA and DoE (USA);
STFC and UKSA (UK); CSIC, MICINN, JA, and RES (Spain); Tekes, AoF, and CSC
(Finland); DLR and MPG (Germany); CSA (Canada); DTU Space (Denmark); SER/SSO
(Switzerland); RCN (Norway); SFI (Ireland); FCT/MCTES (Portugal);
and PRACE (EU). A description of the Planck Collaboration and a list of its
members, including the technical or scientific activities in which they have
been involved, can be found at
\url{http://www.sciops.esa.int/index.php?project=planck&page=Planck_Collaboration}.
Some of the results in this paper have been derived using the {\tt HEALPix}
package.  This research used resources of the National Energy Research
Scientific Computing Center, which is supported by the Office of Science of
the U.S. Department of Energy under Contract No. DE-AC02-05CH11231. 
We acknowledge support from the Science and Technology Facilities Council
[grant number ST/I000976/1].
\end{acknowledgements}

\appendix

\bibliographystyle{aat} 

\bibliography{planck_boosting,Planck_bib}

\raggedright
\end{document}